# Studies on temperature dependent semiconductor to metal transitions in ZnO thin films sparsely doped with Al

Amit K. Das*[1], P. Misra[1], R. S. Ajimsha[1], A. Bose[2], S.C. Joshi[2], D. M. Phase[3] and L. M. Kukreja[1]

[1]Laser Materials Processsing Division, Raja Ramanna Center for Advanced Technology, Indore – 452 013

[2]Proton Linac & Superconducting Cavities Division, Raja Ramanna Center for Advanced Technology, Indore - 452 013

[3]UGC-DAE Council for Scientific Research, Indore – 452 001

*Corresponding author e-mail- amitdas@rrcat.gov.in

**Abstract**

For a detailed study on the semiconductor to metal transition (SMT) in ZnO thin films doped with Al in the concentration range from 0.02 to 2%, we grew these films on (0001) sapphire substrates using sequential pulsed laser deposition. It was found that the Al concentration in the films increased monotonically with the ratio of ablation durations of the Alumina and ZnO targets used during the deposition. Using X-ray photo electron spectroscopy it was found that while most of the Al atoms occupy the Zn sites in the ZnO lattice, a small fraction of the Al also gets into the grain boundaries present in the films. The observed SMT temperature decreased from ~ 270 to ~ 50 K with increase in the Al concentration from 0.02 to 0.25 %. In the Al concentration range of ~ 0.5 to 2 % these doped ZnO films showed metallic behavior at all the temperatures without undergoing any SMT. A theoretical model based on thermal activation of electrons and electron scatterings due to the grain boundaries, ionic impurities and phonons has been developed to explain the observed concentration and temperature dependent SMT.

**I. Introduction**

Studies on the ZnO thin films doped with different elements is an area of contemporary research for the development of new functional materials mainly for photonic and photovoltaic applications[1]. One of the ramifications of this area of research is to obtain and study highly conductive and transparent films of ZnO using n-type dopants such as Al, In, B, Ga and Si[2-7]. Al has been extensively used for this purpose because of its easy availability, low cost, ease of doping and superior properties of the ensuing films. Studies on temperature dependent transport properties of such Al doped ZnO (AZO) films[8, 9] are essential to understand the conduction mechanisms under different conditions of the doping. An important effect that has been observed in the temperature dependent resistivity measurements of AZO thin films is the semiconductor to metal transition (SMT)[10-12]. SMT is an important phenomenon because it can be used to elicit the basic processes involved in the transport properties of AZO films. It is well known that as grown pure ZnO thin films are semiconducting in nature showing decrease of resistivity ($\rho$) with increase of temperature (i.e. it has negative temperature coefficient of resistivity)[2]. When doped with small amounts of (~ 2%) n-type dopant, the ZnO thin films become metallic showing increase of resistivity with increasing temperature (i.e. positive temperature coefficient of resistivity)[6, 10]. Hence it is expected that the temperature dependent resistivity of sparsely doped AZO films with Al concentration < ~ 2% would show SMT. Although SMT in AZO thin films has been reported previously [10-12], a systematic and detailed study of SMT in sparsely doped AZO thin films with controlled doping concentration is lacking. We studied the SMT in sparsely doped AZO thin films as a function of Al doping concentrations and temperature. We also investigated the spatial distribution and valence states of the Al dopant in these AZO films to obtain requisite information for developing a theoretical model to explain the observed SMT in temperature dependent resistivity of the AZO films with different Al concentrations. This theoretical model is based on the donor concentration dependent thermal activation of electrons and temperature dependent electron scattering mechanisms of different kinds involving grain boundaries, ionic impurities and phonons. The results of these studies are presented and discussed in this paper.

## II. Experimental

We used the sequential pulsed laser deposition (SPLD) methodology[7, 13, 14] for growing the AZO thin films used in this study with different Al concentrations in the range from 0.02 to 2% on (0001) sapphire substrates. In this methodology the solid sintered targets of $Al_2O_3$ and ZnO were ablated independently and sequentially for different durations to incorporate the Al dopant with controlled concentrations in the ZnO films. SPLD was necessary to obtain such low concentrations of Al in the AZO thin films with homogeneous spatial distribution of the dopant because the conventional pulsed laser deposition using a single ablation target of Alumina mixed ZnO did not produce homogeneous doping of Al in the ZnO films. To control the concentration of Al in the AZO films during SPLD the $Al_2O_3$ pellet was ablated for one pulse and the number of pulses of ablation of ZnO pellet was varied from 500 to 12 in each cycle of the repetitive ablation, which enabled us to obtain concentrations of Al in the range of 0.02 to 2%. The SPLD was carried out using a KrF excimer laser (248 nm, 20 ns and 10 Hz repetition rate) at a fluence of ~ 0.8 $J/cm^2$. Prior to deposition of the AZO films at $400^0C$, a ZnO buffer layer of thickness of ~ 30 nm was grown on the sapphire substrates at $750^oC$ in each case. The buffer layer was necessary to improve the crystalline quality of the over layers of AZO[13]. The buffer layer and the AZO films were deposited in oxygen ambient at a pressure of $~1.2 \times 10^{-4}$ mbar in a PLD chamber which was initially evacuated to a base pressure of $~5 \times 10^{-6}$ mbar. The total thickness of the over layers of AZO in all the case was ~ 200 nm. The concentrations of Al in the films, including the depth profile, were measured by using a commercial time of flight secondary ion mass spectrometer (TOF-SIMS), with the analysis beam of $Bi^+$ and sputtering beam of $Cs^+$. The Al concentration was calibrated with respect to a standard sample with a known Al concentration. Valence states of the Al in the AZO films were evaluated by XPS measurements using Al $K\alpha$ radiation. High resolution XRD measurements and room temperature photoluminescence from these AZO films were used to ascertain the crystalline quality of the films. The transmission spectra of the films were recorded using a UV-Visible spectrophotometer in the spectral range from 250 to 800 nm. Electrical measurements on the AZO films were carried out in the 4-probe van der Pauw geometry using a lock-in amplifier. For this the electrical contacts were made using Indium, the ohmic nature of which were confirmed from the I-V measurements across the contacts.

## III. Results and discussion

Transmission spectra of all the AZO films showed that their transparency was ~ 85 % in the visible spectral range with a sharp band edge in the UV region. The band edge shifted from ~ 3.28 eV for undoped ZnO to ~ 3.53 eV for ~ 0.50 % Al doped ZnO. This blue shift was found to be due to the Moss-Burstain shift[15, 16]. However this blue shift was partially suppressed due to the competing effect of narrowing of the bandgap resulting from the many body effects [7, 17]. The crystalline quality of the AZO films and their highly c-axis oriented growth in mosaic morphology were confirmed using the normal $\omega$-$2\theta$ and $\omega$-rocking curves of their XRD patterns. Photoluminescence (PL) measurements of these AZO films showed strong excitonic near band edge UV emission without any deep level luminescence. The intensity of the excitonic PL peak decreased with increasing Al doping due to slight deterioration in the crystalline quality of the AZO film[18].

Figure 1 shows the variation of the Al doping concentration as measured by the TOF-SIMS as a function of the ratios of ablation durations of $Al_2O_3$ and ZnO pellets ($T_{Alumina}/T_{ZnO}$). On the right side of this figure the vertical axis shows the corresponding Al concentration in atomic %. It can be seen in this figure that the Al doping concentration in the AZO films varies linearly with $T_{Alumina}/T_{ZnO}$. This shows that sequential PLD can indeed be an effective technique for growing sparsely doped AZO thin films with control on doping concentration. The inset of figure 1 shows the depth profile of Al concentrations measured using TOF-SIMS for two representative films with Al concentration of $4.0 \times 10^{19}$ cm$^{-3}$ (0.10 %) and $2.1 \times 10^{20}$ cm$^{-3}$ (0.5 %). It is clear from these figures that the concentration of Al was nearly constant along the depth of the AZO films up to the thickness of ~ 200 nm. Beyond ~ 200 nm the Al concentration gradually increased plausibly because of diffusion of Al from the sapphire substrate into the ZnO film[19] grown by the SPLD.

Figure 2 shows the XPS profiles of Al 2p peak in the AZO film with Al concentration of ~0.50 %. The Al 2p spectrum was deconvoluted into two component peaks at ~73.8 eV and ~71.8 eV. The peak at 73.8 eV was assigned to Al at Zn substitutional sites in an oxygen deficient ZnO matrix and the lower energy peak at 71.8 eV was assigned to elemental Al in the films, which could be present at the grain boundaries as has been reported earlier[20]. In the AZO films with Al concentrations lower than 0.50 %, we could not detect the Al 2p peaks, understandably due to low values of the ionization cross section of Al. The XPS peaks corresponding to Zn 2p and O1s were also observed conspicuously at their respective standard values of the binding energies.

The results of room temperature electrical measurements are shown in figure 3. As can be seen, with increasing Al concentration from 0 to ~ 0.50 % the resistivity of the films decreased from ~ $4.8 \times 10^{-2}$ to ~ $8.5 \times 10^{-4}$ ohm-cm. The minimum resistivity obtained in this study was ~ $2.8 \times 10^{-4}$ ohm-cm for slightly higher Al concentration of ~ 2.0 %. Carrier concentration increased from $1.4 \times 10^{19}$ cm$^{-3}$ for 0.02 % AZO film to $1.1 \times 10^{20}$ cm$^{-3}$ for 0.50 % AZO film as can be seen in the inset of figure 3. Here it is worthwhile to note that we could control the electron concentration in these AZO films in small steps due to the precise control on the doping level achievable in the SPLD methodology. A comparison of Al concentration measured using TOF SIMS (shown in figure 1) and the electron concentration obtained through Hall measurements (shown in figure 3) revealed that as the Al concentration was increased in the AZO films, the corresponding electron concentration did not increase proportionately. This implies that with increasing concentration of the Al dopant, the fraction of Al atoms that substituted Zn in the ZnO lattice decreased and the fraction which got deposited as elemental Al at the grain boundaries increased. The inset of figure 3 shows that the electron mobility increased from ~ 28 $cm^2$ $V^{-1}$ $s^{-1}$ for the undoped ZnO films to ~ 58 $cm^2$ $V^{-1}$ $s^{-1}$ for the ZnO films doped with ~ 0.50 % Al. This increase in the mobility was attributed to the suppression of the electron scattering due to the grain boundaries present in the AZO films[5, 9]. Beyond an Al concentration of 0.5% the electron mobility mobility decreased which was attributed to the enhanced scattering of electrons with the ionized impurities.

Figure 4 shows the results of the temperature dependent resistivity measurements of the AZO films in the range of 30 to 300 K. The film with Al concentration of 0.02% showed a characteristics semiconducting behavior wherein the resistivity increased with decreasing temperature as shown in the figure 4a. With increase in the doping concentration of Al to 0.05%, the temperature dependent resistivity exhibited a semiconductor to metal transition (SMT) at ~ 270 K indicating that the film was metallic at temperatures higher than ~ 270 K and semiconducting at lower temperatures as shown in fig 4b. The SMT temperature was found to decrease from ~ 270 to ~ 50 K on increasing Al concentration from ~ 0.05 to ~ 0.25%. The SMT eventually vanished for the films with Al concentration of ~ 0.50% and 2%. These films were completely metallic in the entire range of temperature as shown in figure 4g and 4h respectively. To clearly elucidate the observed variation of resistivity, temperature dependent Hall measurements were carried out on the samples selected with the highest, lowest and intermediate dopant concentrations of Al. The measured electron concentration and mobility as a function of temperature for the three selected concentrations of the Al dopant are shown in figure 5. It can be clearly seen in figure 5a that at the Al concentration of 0.02%, the electron concentration increased with the temperature, which was attributed to the activation of electron from the native donors or from the trap states[10, 11]. This is discussed further in our theoretical model presented later. The electron mobility in this case was almost constant throughout the entire measurement temperature range, except near the room temperature, where it decreased slightly. At the Al concentrations of 0.1% as seen in figure 5b, the electron concentration increased sluggishly with temperature compared to that in the previous case, which is understood to be due to the decreased dominance of carrier activation. The electron mobility in this case was nearly constant up to ~ 150 K and beyond that it started decreasing. This behavior was due to the prominence of temperature dependent phonon scattering and suppression of grain boundary scattering, which are discussed in detail later in the theoretical model. At the highest concentration of Al of ~ 0.5% as seen in figure 5c, the electron concentration was conspicuously constant because the carriers being added due to the thermal activation was found to be almost negligible compared to the contribution from the Al dopant. The electron mobility, on the other hand, showed a drastic decrease beyond ~ 100 K primarily due to phonon scattering. From figure 5, it can also be seen that with increase in the Al concentration, the electron mobility increased for the corresponding temperatures, which was attributed to be due to the radical shift in the dominant scattering mechanism from grain boundary to lattice vibrations as discussed in the theoretical model presented later. Some earlier reports in the literature have also shown similar behaviors of electron concentration and mobility in transparent conducting oxide films[21, 22].

Observation of SMT in the temperature dependent resistivity of Ga doped ZnO thin films has been reported earlier by Bhosle *et al*[6]. However, in their case the Ga concentration was much higher (~ 2-5%) than the concentration of Al in the present case and the SMT temperature was found to increase with increasing Ga concentration. In contrast to these observations, our finding is that SMT temperature decreased with increasing concentration of Al in the range of 0.02 to 2%. Bhosle *et al.* attributed the SMT observed by them to the disorder induced weak localization of electrons which increased with increasing dopant concentration, thereby resulting in the increase of transition temperature. We think that this weak localization model is not suitable for our case where Al concentration was sparse (less than 2 %). Further, in previous reports on SMT for Al-doped ZnO[10-12], they qualitatively attributed the positive temperature coefficient of resistance (TCR) above SMT temperature to the formation of degenerated bands and the negative TCR below SMT temperature to the trapping of electrons in the defect states. In our views the SMT in ZnO thin films sparsely doped with Al is not very well established and understood. Here we propose a theoretical model to explain the SMT in AZO thin films based on donor concentration dependent thermal activation of electrons and scattering mechanisms involving grain boundaries, ionic impurities and phonons.

It is a well known fact that in general the metallic behavior is dominated by carrier scattering and the semiconducting behavior is dominated by carrier activation. Since we observed a combination of metallic and semiconducting behavior in the samples it may be possible to explain the observed SMT in our case by a combined mechanism of carrier scattering and activation. This fact is also supported by the observed temperature dependence of electron concentration and mobility as shown in figure 5. The resistivity of the AZO films ($\rho$) can be written as,

$$\rho = \frac{1}{ne\mu} \qquad (1)$$

Where $n$ is the electron concentration, $e$ is the electronic charge and $\mu$ is the electron mobility. So the temperature dependence of $\rho$ is decided by the temperature dependences of $\mu$ as well as that of $n$. Since ZnO thin films with electron concentration higher than $\sim 1 \times 10^{19}$ cm$^{-3}$ are degenerate[23], the functional form of $n$ for Al doped ZnO films can be written as:

$$n = n_0 \left(1 + n_1 \exp(-\phi / kT)\right) \qquad (2)$$

Here we have assumed that the Al dopant atoms in the ZnO films are completely ionized and their concentration $n_0$ is independent of temperature[23]. We attribute the second term corresponding to the carrier activation to be due to the native donors in ZnO or activation of electrons from trap states[10, 11] with the activation energy of $\phi$, which is also a fitting parameter in our calculations. $T$ is the absolute temperature, $k$ is the Boltzmann's constant and $n_1$ is the additional carrier concentration contributed to $n$ due to the thermal activation.

The electron mobility in the present case is determined by various scattering mechanisms. In polycrystalline n-type doped ZnO thin films the primary mechanisms of electron scatterings are due to the grain boundaries, ionized impurities and lattice vibrations. The total mobility in this scenario is given by[21]:

$$1/\mu = 1/\mu_g + 1/\mu_i + 1/\mu_{ph} \quad (3)$$

Where $\mu_g$, $\mu_i$ and $\mu_{ph}$ are the contributions to the mobility due to scatterings from grain boundaries, ionized impurities and lattice vibrations respectively. Grain boundary scattering arises due to the presence of potential barriers for electrons at the grain boundaries. Electrons can either be thermally emitted across the grain boundaries or they could tunnel through. Out of these two mechanisms the dominant one is decided by the tunneling parameter denoted as $E_{00}$ and given by Crowell et al.[21, 24] in the following form:

$$E_{00} = 18.5 \times 10^{-12} \left( \frac{n}{m^* \varepsilon} \right)^{1/2} \text{eV} \quad (4)$$

Where $\varepsilon$ is the relative static dielectric constant, $m^*$ is the electron effective mass and $n$ is the electron concentration as stated earlier in equation 2. For ZnO, $\varepsilon = 8.65$ and $m^* = 0.28 m_e$[7] where $m_e$ is the free electron mass. If $E_{00}$ is larger than $kT$, tunneling will be the dominant mechanism of electron conduction across the grain boundaries. For $n = 1.4 \times 10^{19}$ cm$^{-3}$, the value of $E_{00}$ is ~ 45 meV which is greater than $kT$ in the range of the measurement temperatures used in this study. It is therefore expected that the current will be dominated by tunneling through the grain boundaries rather than thermionic emission[21]. Since tunneling is independent of temperature[21, 25], the mobility due to grain boundary scattering ($\mu_g$) can also be taken to be independent of temperature. In ZnO with an electron concentration of > $1 \times 10^{19}$ cm$^{-3}$, the Fermi level is expected to be in the conduction band[23]. In such a case the ionized impurity scattering is also expected to be independent of temperature[21]. Hence mobility due to ionized impurity scattering ($\mu_i$) is taken to be constant. The lattice vibration scattering for the degenerate semiconductor such as ZnO in the present case can be written similar to that of metals[21] with $\mu_{ph} \propto T^{-p}$, where $p$ is 1 if $T$ is greater than the Debye temperature[26] and it ranges between 2 and 4 if $T$ is less than the Debye temperature[27]. Since for ZnO the Debye temperature is ~ 400 K, which is much higher than our measurement temperatures, we took a value for $p$ as 2 after Nistor et al[28]. Combining all the scattering mechanisms, the overall mobility can be given by,

$$1/\mu = C + DT^2 \quad (5)$$

Where $C$ and $D$ are constants. It can be seen from the figure 5, that the Hall mobilities of the films could be described by equation (5) with appropriate values of the parameters C and D (fitting not shown). Combining equations (1), (2) and (5) and after simplification we get following relation for the resistivity $\rho$:

$$\rho = \frac{A(1+BT^2)}{1+n_1 \exp(-\phi/kT)} \quad (6)$$

Where $A$, $B$, $n_1$ are temperature independent constants and taken as fitting parameters. The calculated curves obtained from equation (6) with the experimental data points are shown in the figure 3. As can be seen a very good fit with the experimental data is obtained using this equation. The values of various fitting parameters for the samples are listed in Table 1. The value of the fitting parameter $n_1$ gradually decreases with increasing Al concentration i.e., free electron concentration. It implies that the contribution of carrier activation is gradually decreasing with increasing doping concentration. This is also supported by the fact that the corresponding activation energy $\phi$ decreases with the increasing concentration of Al, which can be attributed to the increasing screening of the Coulomb potential of the donors with increasing electron concentration[26] and/or band tailing arising due to heavy doping[23].

The decrease in the SMT temperature in sparsely doped AZO thin films with increasing Al concentration can also be qualitatively explained by equation (6). As Al concentration increases i.e. electron concentration increases, the activation energy of the native donors or trap states decreases. This is reflected by the decrease in the fitting parameters $n_1$ and $\phi$. Hence the contribution of carrier activation behavior in total resistivity decreases as compared to the contribution of carrier scatterings, thereby reducing the temperature at which the SMT occurs. In the AZO films with Al concentration at or above 0.50 %, the carrier activation is almost ruled out and resistivity is dominated by carrier scatterings in the entire range of measurement temperature, which is typical of metallic behavior. The fitting parameter $A$ gives a measure of temperature independent resistivity or residual resistivity of the samples. The fitting parameter $B$ decides the dominance of the phonon scattering with respect to the impurity and grain boundary scatterings. It can be seen from table 1 that the value of $B$ increased with increasing Al doping, indicating enhancement in the contribution of phonon scattering to the overall mobility. This fact is supported by the results of the temperature dependent Hall mobility measurements as shown in figure 5. The reason for the observed enhancement in the contribution of phonon scattering with increasing doping is possibly the gradual change in dominant scattering mechanisms from inter-grain to intra-grain [4, 9] ones. In relatively lightly doped samples the mobility is primarily limited by tunneling mediated grain boundary scattering. As the Al concentration is increased the inter-grain scattering is suppressed and intra-grain scattering mechanisms like phonon scattering start to dominate the overall mobility[4, 9]. Because of the same reason of diminished inter-grain scattering, the mobility in AZO thin films increased on increasing Al doping concentration up to ~ 0.5%, as shown in the inset of figure 3.

**IV. Conclusion**

In conclusion, we have grown high quality ZnO thin films sparsely doped with Al of varying concentrations in the range of 0.02 to 2 % by sequential pulsed laser deposition technique and studied their semiconductor to metal transition (SMT) in detail. SMT was observed in the temperature dependent resistivity measurements in the samples with Al concentration below 0.50 %. SMT transition temperature in AZO films was found to decrease with increasing Al concentration. The 0.5 % and 2% Al doped AZO film showed metallic behavior at all the temperature range without any SMT. The SMT and the associated variation of SMT transition temperature with Al concentration was explained on the basis of a theoretical model based on the competition between donor concentration dependent carrier activation behavior and various scattering mechanisms.

**Acknowledgement:** Authors thank Mr. A. Wadikar of UGC-DAE CSR, Indore for his help in XPS measurements. A part of this study was carried out at UGC-DAE CSR, Indore.

[26]N. F. Mott, *Metal-Insulator Transition* (Taylor and Francis, London, 1974).

[27]X. D. Liu, E. Y. Jiang, and Z. Q. Li, J. Appl. Phys. **102**, 073708 (2007).

[28]M. Nistor, F. Gherendi, N. B. Mandache, C. Hebert, J. Perrière, and W. Seiler, J. Appl. Phys. **106**, 103710 (2009).

**Figure Caption:**

**Figure 1:** Variation of Al concentration in the AZO films as measured by TOF-SIMS with the ablation duration ratios of Alumina and ZnO pellets ($T_{Alumina}/T_{ZnO}$). The inset shows typical SIMS depth profile of Al concentration for the samples with Al concentration of 0.1% and 0.5%.

**Figure 2:** XPS spectrum of the Al 2p peak for the film with Al concentration of 0.50 %. The deconvolution components are shown by the dashed curves whereas the continuous curve shows the fitted data.

**Figure 3:** Variation of resistivity of the AZO films grown by sequential PLD as a function of Al doping concentrations. The inset shows the corresponding variation of electron concentration by filled squares and mobility by filled circles.

**Figure 4:** Temperature dependent resistivity (filled circles) of the AZO films with Al doping concentrations of (a) 0.02%, (b) 0.05%, (c) 0.06%, (d) 0.10%, (e) 0.12%, (f) 0.25%, (g) 0.5% and (h) 2%. The continuous curves show the corresponding fittings of the data according to equation (6).

**Figure 5:** Temperature dependence of electron concentration and Hall mobility for AZO films with Al doping concentrations of (a) 0.02%, (b) 0.10% and (c) 0.5%.

Table I: Values of different parameters of equation (6) obtained from fitting of the temperature dependent resistivity data.

| Al concentration (%) | $A$ | $B$ | $n_1$ | $\phi$ (meV) |
|---|---|---|---|---|
| 0.02 | $(1.65\pm.007)\times10^{-2}$ | $(3.8\pm.1)\times10^{-7}$ | $1.0\pm.1$ | $20.2\pm.5$ |
| 0.05 | $(7.560\pm.007)\times10^{-3}$ | $(6.5\pm0.6)\times10^{-7}$ | $0.34\pm.02$ | $18.2\pm.5$ |
| 0.06 | $(5.720\pm.003)\times10^{-3}$ | $(1.18\pm.03)\times10^{-6}$ | $0.29\pm.01$ | $18.3\pm.3$ |
| 0.10 | $(4.250\pm.003)\times10^{-3}$ | $(1.70\pm.05)\times10^{-6}$ | $0.23\pm.01$ | $17.7\pm.5$ |
| 0.12 | $(1.970\pm.007)\times10^{-3}$ | $(1.82\pm.02)\times10^{-6}$ | $0.15\pm.01$ | $15.8\pm.3$ |
| 0.25 | $(1.600\pm.001)\times10^{-3}$ | $(2.12\pm.06)\times10^{-6}$ | $0.10\pm.01$ | $14.8\pm.9$ |
| 0.50 | $(7.200\pm.005)\times10^{-4}$ | $(3.00\pm.04)\times10^{-6}$ | $0.10\pm.01$ | $12.6\pm.7$ |
| 2.00 | $(2.42\pm.002)\times10^{-4}$ | $(2.76\pm.03)\times10^{-6}$ | $.09\pm.01$ | $13.0\pm.2$ |

Figure 1:

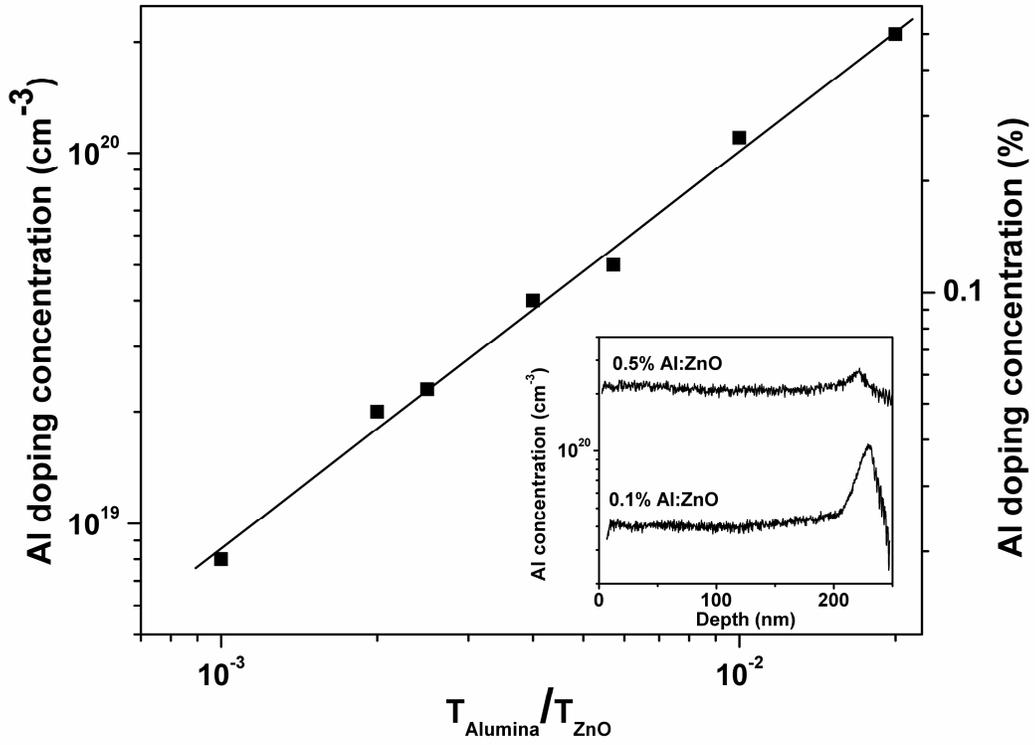

**Figure 2:**

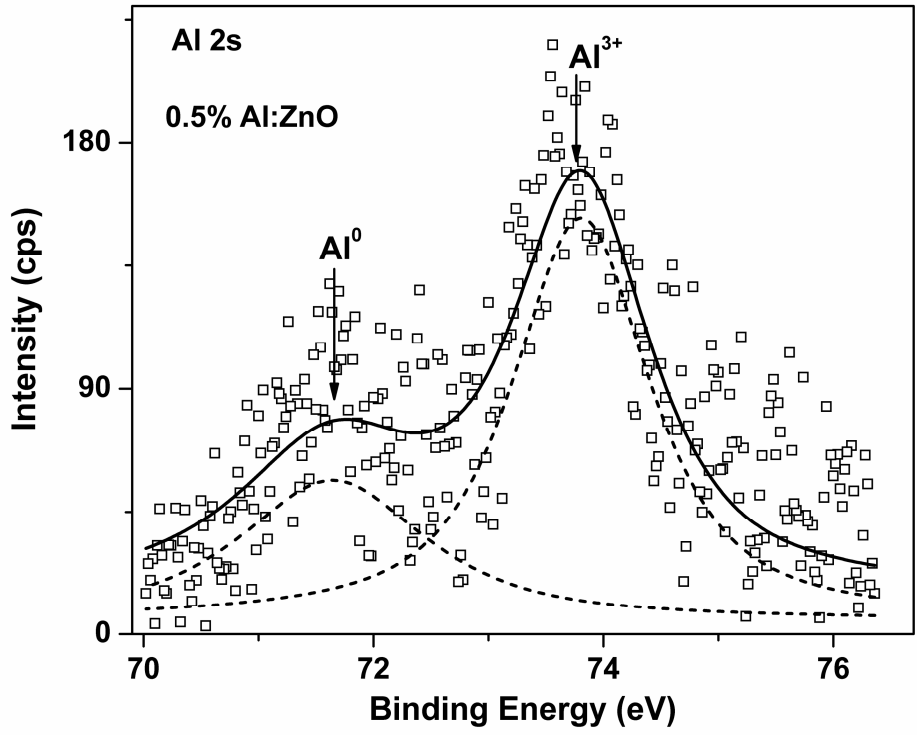

Figure 3:

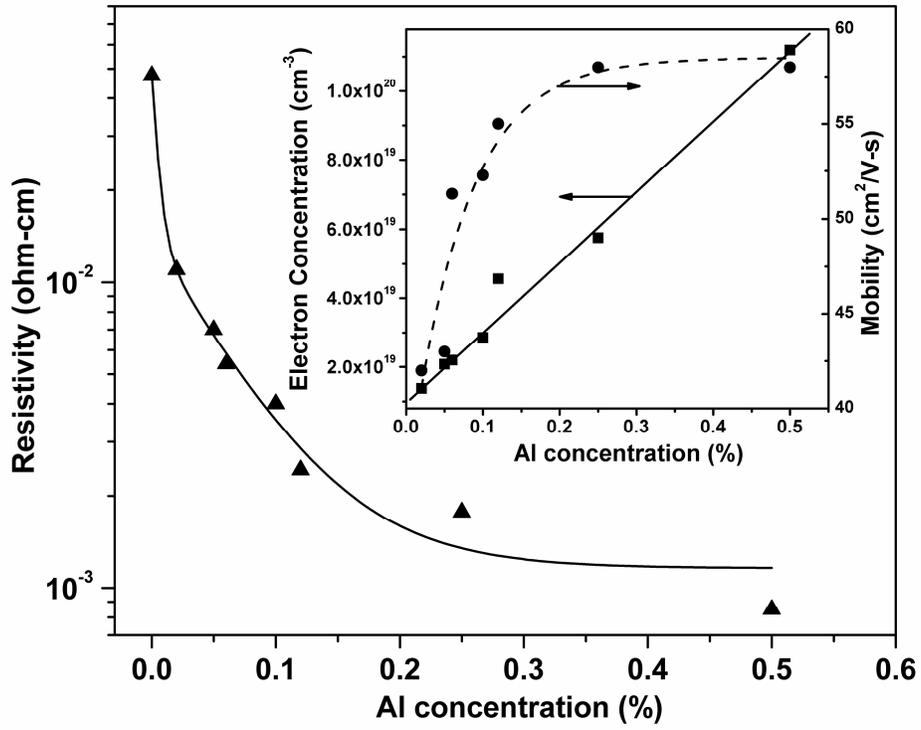

Figure 4:

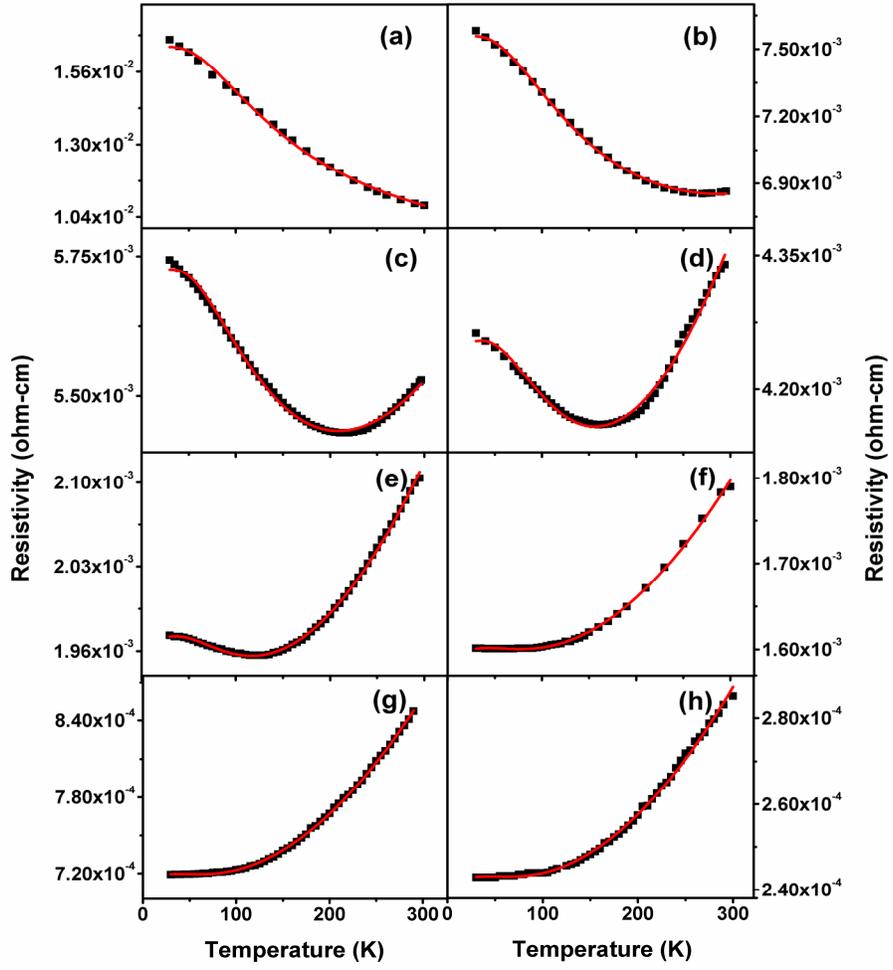

Figure 5.

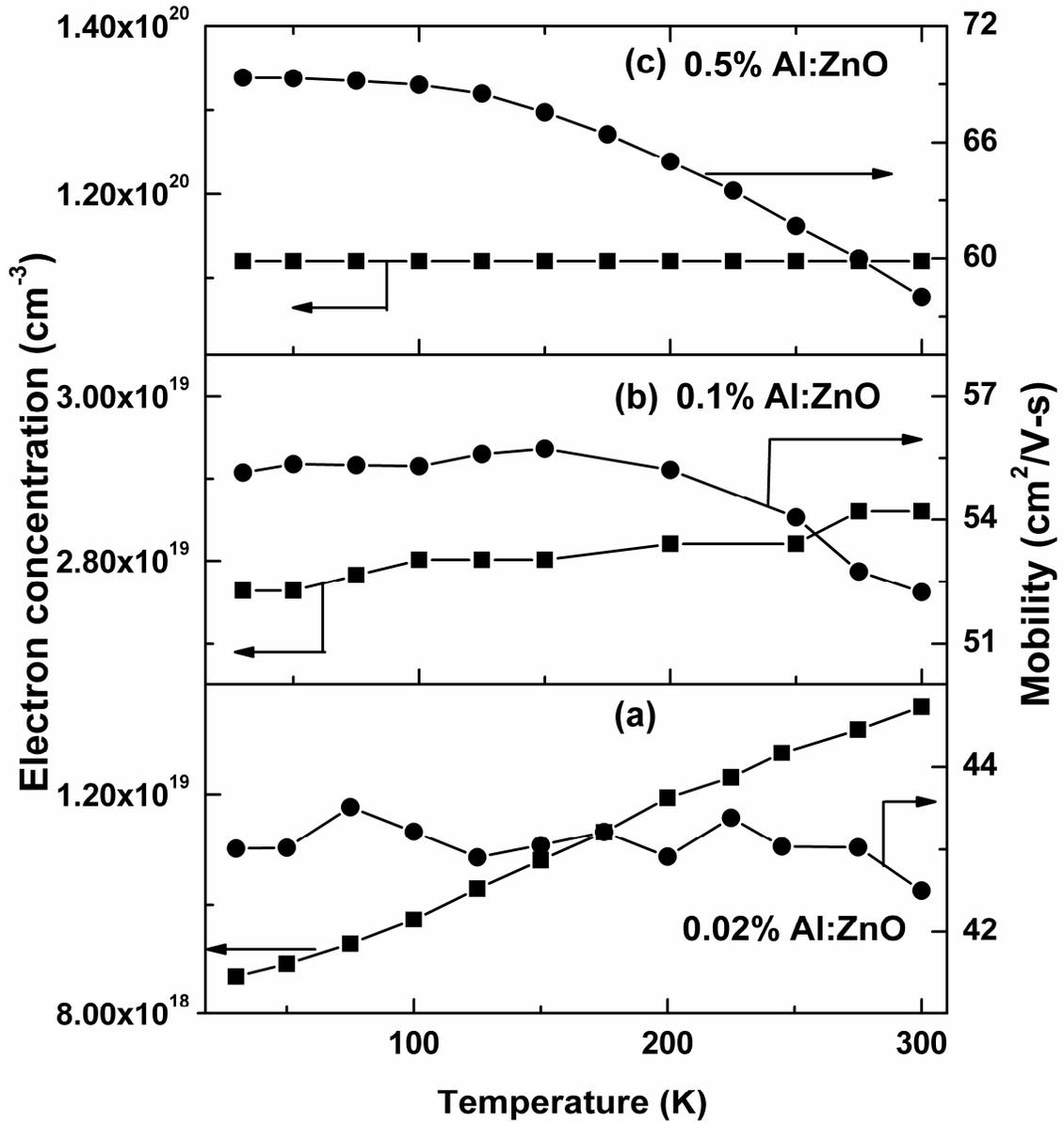